\documentclass[12pt,a4]{article}

\usepackage{tabularx}
\usepackage{graphicx}

\title{Compositional analyses
 of a Reutlingen Bronze Age sword discovered at Giurgiu, Romania}

\author{Agata Olariu$^{\diamond}$, Emilian Alexandrescu$^{\bullet}$,  Alexandru Avram$^{\bullet}$, Teodor Badica$^{\diamond}$\\
{\em $^{\diamond}$Institute for Physics and Nuclear Engineering,}\\
{\em PO Box MG-6, 76900 Magurele, Bucharest, Romania}\\
{\em $^{\bullet}$Institute of Archaeology of the Romanian Academy,}\\
{\em Henri Coand\u a Street, no. 11,  Bucharest, Romania}\\
}
\date{}
\begin{document}
\maketitle
\thispagestyle{empty}
\begin{abstract}
The compositional scheme of a Bronze Age sword, found near the town of Giurgiu in Romania 
has been determined by the method of particle-induced X-ray emission (PIXE), at the Tandem accelerator of National Institute for Physics  and Nuclear Engineering  from Bucharest, Magurele, Romania. The results of the analyses and the comparison with the composition of other swords from the same geographic area, the Danubian plane from Bulgaria and Transylvania regions, show that the sword from Giurgiu could be relatively associated with the swords from Bulgaria, having also the same stylistic, temporal and geographical similitude.

\end{abstract}

\section{Introduction}
The compositional scheme of a Bronze Age sword, Fig. 1, recently discovered near Giurgiu,  a town situated in the south of Romania on the Danube, has been studied using the method of particle-induced X ray emission (PIXE), at the Tandem accelerator of National Institute for Physics  and Nuclear Engineering  from Bucharest, Magurele, Romania.
In order to have a comparative study of composition we have considered also the composition of 6 swords of the same type and from the same period from the south of Danube in Bulgaria, and some different copper-based alloy objects from \c Sp\u alnaca deposit, in Transylvania, dated also in the Bronze Age.

\section{Archaeological considerations}

\subsection{Site of discovery of the sword}

The sword has been discovered by Ion Cercel in 1981 in the Mihai Eminescu street of 
the city of Giurgiu, Romania, in the course of diggings related to the installation
of underground electric cables. Since then the sword has been kept by Ion Cercel, who
presently is retired, in his house from the Oinacu commune, district of Giurgiu, in 
southern Romania. In 1999, by a happy concurrence of events, this interesting artifact
became available to us, and we take this opportunity to thank Ion Cercel who has been kind
to render us the sword for study and analyses.

According to the author of the finding, the artifact has been found at a depth of 
approximately 1.1 m, no other bronze objects existing in its vicinity. Foundations of buildings
from the late medieval period were known to exist in the area, but it seems that no such 
archaeological remains existed in the specific place where the sword has been found. It seems 
that the site where the sword has been discovered lies outside the medieval city, buildings
being erected there only in the period after 1821.

\subsection{Description of the sword}

The sword is in a good state of conservation, being preserved almost entirely, and has the 
following dimensions: total length 39.5 cm, width at base 5.3 cm, width at the tip  2.9 cm,
thickness at base 0.9 cm, thickness toward the tip 0.6 cm, length of 
the hilt with missing terminal part 9 cm, length of the blade 30.5 cm. On the blade of the 
sword one can observe, on both faces, channels situated at approximately 0.4 cm from the 
two edges, as shown in Fig. 1. The blade has a biconvex profile. In the zone of the hilt
there are five orifices for the rivets of binding of the hilt and also three rivets still
left in the orifices. 
It is possible that in the missing part of the hilt three more orifices should have existed
for the binding of the hilt. The alloy is of a very good quality having a green-dark grey patina.

The sword belongs to the Reutlingen type defined by P. Schauer \cite{1} and described, with
a special view to artifacts attested on Romanian territory, by T. Bader. \cite{2}
Choosing as a criterion of classification in the first place the number of fixation orifices
from the blade and the hilt, but also the shape of the nervure of the blade, the latter 
author distinguished several variants. \cite{3} Due to the fact that the sword under 
consideration presents a large median nervure and which is slightly rounded, we think that it
is most resembling to two fragmentary pieces belonging to the deposit of Drajna de Jos, district
of Prahova, Romania, catalogued by T. Bader with numbers 188 and 189 \cite{4} and included in the
Gu\c steri\c ta variant of the Reutlingen type. 
Moreover, the Giurgiu sword has much smaller dimensions, so that it could be rather considered 
a "short sword" (Kurzschwert). \cite{5}

Artifacts of the Reutlingen type have been discovered over a very large area from the south of
Scandinavia to Peloponesos and from the Rhine basin to the Black Sea, \cite{6} and recently
discoveries have been reported even in Anatolia. \cite{7} However, the spreading is not uniform,
existing some regions of concentration and others represented by much fewer discoveries.
Among the latter, one counts the extra-Carpathian zone in Romania and the territory of Bulgaria
and Greece. \cite{8}

It is however interesting that, in the Balkan area, the discoveries are concentrated mainly
in the southern part of Romania, Wallachia,
and in the north of Bulgaria, some of them just on the Danube line. In addition to the two 
pieces from Drajna de Jos, on the Romanian territory one finds specimens belonging to some 
variants in the ensemble of the Reutlingen type: B\u alce\c sti and Matee\c sti (district
of V\^ alcea), Techirghiol (district of Constan\c ta). On the territory of Bulgaria  there are
10 discoveries of swords with tongues at the hilt, 
of which 7 to the north and 3 to the south of the Balkans. \cite{9} The 7 specimens discovered on the
territory between the Danube and the Balkans arise from Orjahovo (Orehovo), \cite{10}
V\u arbica (deposit II), \cite{11}, Bajkal, \cite{12} Kru\v sevo, \cite{13} Balkanski \cite{14}
and Vasil Levski, \cite{15} to which one adds the specimen of smaller dimensions from the 
Razgrad Museum (inventory No. 117), discovered in the neighborhood. \cite{16} Among these specimens the first two belong to the Reutlingen type.

The artifact from Giurgiu has very close analogues (except, of course, for the dimensions) just 
in the sword from Orjahovo and in the fragmentary artifact from V\u arbica, both cited as 
belonging to the Gu\c steri\c ta variant by T. Bader. \cite{17} These two specimens have been 
ascribed in the early horizon of the culture of the fields of urns
(von Brunn stages I-III) \cite{18} by B. H\" ansel, \cite {19} respectively in the subgroup
I defined by  I. Panayotov (the second horizon of deposits from Bulgaria: XIII$^{th}$ century B.C.).
\cite{20} On the other hand, T. Bader dates, as a function of the synchronisms revealed by the
various deposits, the great majority of the specimens belonging to the Reutlingen type discovered
on the territory of Romania in the Cincu-Suseni period (HaA1, circa XII$^{th}$ century B.C.), but
ascribes three or four deposits (among which is also the one from Drajna de Jos) for the
slightly earlier period Uriu-Dom\u ane\c sti (Bronze D, circa XIII$^{th}$ century B.C.) \cite{21}
Consequently, taking into account the analogies proposed by us with the specimens from 
Drajna de Jos, Orjahovo and V\u arbica II, we favor a dating of the short sword from Giurgiu
in the XIII$^{th}$ century B.C., probably towards the end of the century; a date around 1200 B.C. is
very likely.

\section{Experimental}
3 samples from the body of the sword: 1 sample from the tip of the sword and 2 samples from the hilt have been flatted and irradiated with protons of 3 MeV, in a irradiation chamber at  the FN Tandem accelerator of National Institute for Physics  and Nuclear Engineering  from Bucharest, Magurele. 

The beam current was
kept below 10 nA to maintain a count rate of about 250 counts/s, which implies
negligible dead-time and pile-up corrections. X-rays were detected with a HPGe
(100 mm$^2\time10$mm) detector with 160 eV energy resolution at 5.9 keV.
The X rays spectra have been recorded on a PC with a MCA interface.
In the frame of the experimental conditions  the following elements have been observed: As, Co, Cr, Cu, Fe, Ni, Pb, Sn and Zn.

The X ray spectra have been processed off line and then the concentrations of the elements have been calculated.

\section{Results and Discussions}
The results of PIXE analysis on the samples from the sword from Giurgiu are shown in the Table 1. The values of the concentrations are given in \%. The instrumental errors are generally less than 15 \%.
We made corrections of the elemental concentrations so that the total value in the sample to be 100 \%.

\begin{center}

\begin{table}

\caption{Composition of the sword from Giurgiu, by PIXE}

\begin{center}

\begin{tabular}{lrrrrrrr}
\hline
Sample  &    As  &  Co    &   Cu   &     Fe   &    Ni   &    Sn   &   Zn    \\
\hline
\hline
Sword tip  &    0.3530  &   0.0440    &     88.2  &   0.0838   &   0.3090  &10.4     &    0.6173   \\
Sword  big hilt     &    0.0855   &     0.0171    &     85.5     &    0.3850   &       0.3250     &    13.7     &    0.0470  \\
Sword small hilt       &    0.2860   &     0.0224   &      89.5     &    0.4740    &      0.3400     &     9.35    &    0.0313  \\
          
\hline
\end{tabular}
\end{center}
\end{table}
\end{center}
The composition of the 3 samples form the Giurgiu has been compared with the composition from similar 6 swords from Danubian regions from Bulgaria \cite{22} and some different archaeological objects from the Bronze Age \c Sp\u alnaca deposit, Transylvania \cite{23}.

We present further, in the Table 2 the results of the analyses published by E. N. \v Cernyh, for several of the swords with tongue at the hilt from Bulgaria. \cite{22}\\

\noindent
1: V\u arbica II (10945), category X\\
2: Orjahovo (9431), category X\\
3: Pavelsko (9220), category X\\
4: Bajkal (9432, analysis of the hilt; 9433, analysis of the blade), category X\\
5: Kri\v cim (9210), category X\\
6: Vasil Levski (10892), category XI\\

\noindent
For all specimens included in Table 2, the copper is the dominant element.

\begin{table}

\caption{Composition of swords from Bulgaria, \% \cite{22}}

\begin{tabular}{lrrrrrrrrrrrr}
\hline
 &   Sn  &  Pb  & Zn &  Bi  &  Ag &  Sb  & As & Fe & Ni & Co & Mn & Au   \\
\hline
\hline
1 & 10 & 0.2 & 0.01 & 0.05 &  0.06 & 0.06 & 0.07 &0.007& 0.05 & 0.02 &-& $<$0.001\\
2 & 10 & 0.14& 0.01 & 0.01 & 0.06 & 0.04 & 0.6 & 0.05& 0.4& 0.04 & - & $\approx$0.01\\
3 & 12 & 0.3 & ? &    0.005 & 0.06 & 0.25 & 0.3& 0.003& 0.25 & 0.03 &- &$\approx$0.003\\
4 & 7 & 0.12 & 0.006& 0.003 & 0.05 & 0.3 & 0.8& 0.01& 0.3 & 0.012 &- &$<$0.01\\
  & 10 & 0.3 & ? &    0.005 & 0.03 & 0.3 & 0.9 & ? & 0.35 & 0.02 & - &$>$0.001\\
5 & 7 & 0.05 & - &    0.0015 & 0.01 & 0.04 &0.25 & 0.005 & 0.05 & 0.015&$<0.01$&$>$0.003\\ 
6 & 5 & 0.09 & - &    -     & 0.0001& 0.015 & 0.1 & 0.012 & 0.035 & 0.003 & -&-\\
\hline
\end{tabular}
\end{table}

\begin{table}
\caption{Ratios of concentrations, in bronze objects of the same type: Giurgiu sword samples, 
by PIXE, the Bulgarian swords, by atomic spectroscopy, bronze objects from 
Splanaca, Transylvania, by neutron activation}
\begin{center}
\footnotesize
\begin{tabular}{lrrrrrr}
\hline
Sample
&
As/Cu
&
Co/Cu
&
Fe/Cu
&
Ni/Cu
&
Sn/Cu
&
Zn/Cu\\
& x 10$^6$& x 10$^6$& x 10$^6$& x 10$^6$& x 10$^6$& x 10$^6$
\\
\hline
\hline
Giurgiu1
&
4000
&
500
&
950
&
3500
&
118000
&
7000
\\
Giurgiu2
&
1000
&
200
&
4500
&
3800
&
160000
&
550
\\
Giurgiu3
&
3200
&
250
&
5300
&
3800
&
104500
&
350
\\
Bulgaria1
&
779
&
223&
78&
556&
111305
&
111\\
Bulgaria2
&
6750&
450&
562&
4500
&
112500
&
112\\
Bulgaria3
&
3430&
343&
34.3&
2860
&
137300
&
0
\\
Bulgaria4
&
8710
&
131
&
109
&
3265
&
76190
&
65\\
Bulgaria5
&
10140&
225&
0
&
3945
&
112700
&
0
\\
Bulgaria6
&
2700&
162&
53.9&
539&
75530
&
0
\\
Bulgaria7
&
1055&
32&
1270&
369.4
&
52780
&
0
\\
\c Sp\u alnaca1
&
6848
&
0
&
0
&
0
&
188600
&
0
\\
\c Sp\u alnaca2
&
1193
&
0
&
79700
&
0
&
3250
&
0
\\
\c Sp\u alnaca3
&
16600
&
0
&
44400
&
0
&
0
&
0
\\
\c Sp\u alnaca4
&
12300
&
0
&
50900
&
0
&
0
&
0
\\
\c Sp\u alnaca5
&
13100
&
0
&
0
&
0
&
81600
&
0
\\
\c Sp\u alnaca6
&
22100
&
0
&
0
&
0
&
0
&
0
\\
\c Sp\u alnaca7
&
15900
&
0
&
135000
&
0
&
0
&
0
\\
\c Sp\u alnaca8
&
23000
&
0
&
32900
&
0
&
0
&
0
\\
\c Sp\u alnaca9
&
67400
&
0
&
98100
&
0
&
0
&
0
\\
\c Sp\u alnaca10
&
2090
&
0
&
0
&
0
&
0
&
0
\\
\c Sp\u alnaca11
&
7900
&
0
&
21600
&
0
&
0
&
0
\\
\c Sp\u alnaca12
&
7180
&
0
&
0
&
0
&
203100
&
0
\\
\c Sp\u alnaca13
&
15600
&
0
&
12600
&
0
&
1770
&
0
\\
\c Sp\u alnaca14
&
10360
&
0
&
334000
&
0
&
0
&
0
\\
\c Sp\u alnaca15
&
2408
&
0
&
0
&
0
&
0
&
0
\\
\c Sp\u alnaca16
&
13170
&
0
&
7970
&
0
&
0
&
0
\\
\c Sp\u alnaca17
&
7460
&
0
&
49700
&
0
&
0
&
0
\\
\c Sp\u alnaca18
&
2900
&
0
&
0
&
0
&
253400
&
0
\\
\c Sp\u alnaca19
&
3400
&
0
&
0
&
0
&
0
&
0
\\
\c Sp\u alnaca20
&
7160
&
0
&
0
&
0
&
0
&
0
\\
\c Sp\u alnaca21
&
44900
&
0
&
0
&
0
&
0
&
0
\\
\c Sp\u alnaca22
&
19200
&
0
&
0
&
0
&
0
&
0
\\
\c Sp\u alnaca23
&
32600
&
0
&
19800
&
0
&
4200
&
0
\\
\c Sp\u alnaca24
&
8900
&
0
&
0
&
0
&
0
&
0
\\
\hline
\end{tabular}
\end{center}
\end{table}

In Table 3 are shown the elemental composition for all considered objects: the Giurgiu sword, the swords from Bulgaria and different bronze objects from Transylvanian deposit at \c Sp\u alnaca. Ratios of concentrations are considered for interpretation of the results to avoid the errors in the absolute calculations of the concentrations. It has been reported value zero in the cases the value of concentrations has been under the limit of detection.
 
Fig. 2  presents the diagram of ratios of concentrations: Sn/Cu versus As/Sn for the analyzed samples in the present study, and also for Bulgarian and Transylvanian objects, analyzed by atomic spectroscopy and respectively neutron activation analysis.

One could remark that the sword from Giurgiu has a relative closer composition to the Bulgarian ones, especially for the elements: As, Cu, and Sn. 
The objects from Transylvania are situated relatively outside the cluster formed by the objects from Giurgiu and Bulgaria.

\clearpage

\section{Conclusions}
We could express the idea of an association of the sword from Giurgiu with the Bulgarian swords, having a close composition and also similitude in typology, geographic area and dating.
Taking into account the analogies proposed by us with the Bulgarian specimens,
Especially those of
Drajna de Jos, Orjahovo and V\u arbica II, we favor a dating of the short sword from Giurgiu
in the XIII$^{th}$ century B.C., probably towards the end of that century, around 1200 B.C.


\begin{thebibliography}{20}
\bibitem{19} B. H\" ansel, op. cit., p. 34 and note 19, respectively p. 35, where the 
fragment of sword from V\u arbica is dated in conjunction with other constituent pieces 
of the deposit which present analogies with types from Gu\c steri\c ta: "die \" ubrigen 
Gegenst\" ande des Hortfundes legen seine Datierung in die fr\" uhe Urnenfelderzeit, d.h.
in die v. Brunnschen Stufen I und II mit hinl\" anglicher Sicherheit fest". Cf. p. 36: "beide
[the items from Orjahovo and V\u arbica (authors' note)] vertreten den gleichen Horizont bzw.
die $<<$Typenfront$>>$, die in Griechenland  mit dem Ende der mykenischen Zivilisation im 
sp\" aten 13. Jahrhundert verkn\" upft ist".
\end{thebibliography}
\end{document}